\documentclass[rnote]{aa}
\usepackage{graphicx}
\usepackage{txfonts}

\newcommand{\MSOL}{\mbox{$M_{\sun}$}}

\newcommand{\micron}{\mbox{$\mu$m}}
\newcommand{\KMS}{\mbox{km s$^{-1}$}}
\newcommand{\HOH}{\mbox{H$_2$O}}

\newcommand{\VLSR}{\mbox{$V_{\rm LSR}$}}
\begin{document}
\title{
VLA survey of the 22~GHz \HOH\ masers toward \\
10 silicate carbon stars 
\thanks{
Based on MIDI observations made with the Very Large Telescope 
Interferometer of the European Southern Observatory. 
Program ID: 076.D-0250
}
\fnmsep
\thanks{
This work is based [in part] on observations made with the Spitzer 
Space Telescope, which is operated by the Jet Propulsion Laboratory, 
California Institute of Technology under a contract with NASA.
}
}
%\subtitle{
%}

\author{K.~Ohnaka\inst{1} 
\and
D.~A.~Boboltz\inst{2,3}
\and
G.~Mulitz-Schimel\inst{2}
\and
H.~Izumiura\inst{4}
\and
M.~Wittkowski\inst{5}
}

\offprints{K.~Ohnaka}

\institute{
Max-Planck-Institut f\"{u}r Radioastronomie, 
Auf dem H\"{u}gel 69, 53121 Bonn, Germany\\
\email{kohnaka@mpifr.de}
\and
United States Naval Observatory, 3450 Massachusetts Avenue, NW, 
Washington, DC 20392-5420, U.S.A.
\and
National Science Foundation, 4201 Wilson Boulevard, Arlington, VA, 22230, USA \\
\email{dboboltz@nsf.gov}
\and
Okayama Astrophysical Observatory, National Astronomical Observatory, 
Kamogata, Asakuchi, Okayama 719-0232, Japan
\and
European Southern Observatory, Karl-Schwarzschild-Str.~2, 
85748 Garching, Germany
}

\date{Received / Accepted }

\abstract
{
Despite their carbon-rich photospheres, 
silicate carbon stars show evidence of oxygen-rich circumstellar material, 
which is considered to exist in disks.  
Silicate carbon stars represent interesting cases that allow us 
to study the possible effects of binarity on stellar evolution and the mass loss 
accompanied by the formation of disks.  
}
% Aim
{
We present a small survey of the 22~GHz \HOH\ masers toward 10 silicate 
carbon stars with much better sensitivity than the previous observations.  
}
% Methods
{
We observed our sample with the Karl G. Jansky Very Large Array (VLA) using the 
most expanded configuration (A-configuration) with a maximum baseline 
of 36~km. 
For some of our program stars with noisy IRAS Low Resolution Spectra 
(LRS), we present new mid-IR spectra obtained with the Very 
Large Telescope Interferometer and the Spitzer Space Telescope. 
}
% Results
{
We detected the \HOH\ masers toward 5 out of 10 silicate carbon 
stars (EU~And, V778~Cyg, IRAS06017+1011, V1415~Cyg, and NC83=V1945~Cyg), 
with NC83 being new detection.  
No \HOH\ masers were detected toward BM~Gem, IRAS07221-0431, IRAS08002-3803, 
IRAS18006-3213, and HD189605.  
The velocity separation between the most blue- and red-shifted maser features 
is 10--14~\KMS.
If we assume that the masers originate in circum-companion disks, 
the measured velocity separations translate into a lower limit of 
the rotational velocity of 5--7~\KMS, and the upper limit of 
the radius of the maser emitting region is estimated 
to be 10--68~AU for a companion mass of 0.5--1.7~\MSOL.  
The new mid-IR spectra of NC83, IRAS06017+1011, and HD189605 
confirm the 10~\micron\ silicate emission.  The latter two stars show a bump 
at $\sim$11.5~\micron, which is presumably due to SiC originating in 
the ongoing mass loss from the carbon-rich primary star, not due to 
crystalline silicate.  
We also report on the detection of the UV flux at 2271~\AA\ toward HD189605. 
}
% Conclusions
{}

   %  END OF ABSTRACT

\titlerunning{
VLA survey of \HOH\ masers toward 10 silicate carbon stars
}
\authorrunning{Ohnaka et al.}
\maketitle

\begin{table*}
\begin{center}
\caption {Our program stars.  
Y: detection (new detection marked in boldface), N: non-detection. 
References: 
BLM87: Benson \& Little-Marenin (\cite{benson87}). 
BLM96: Benson \& Little-Marenin (\cite{benson96}). 
C00: Colomer et al. (\cite{colomer00}).  
D88: Deguchi et al. (\cite{deguchi88}). 
D89: Deguchi et al. (\cite{deguchi89}). 
E94: Engels (\cite{engels94}). 
EL94: Engels \& Leinert (\cite{engels&leinert94}). 
EPC: Engels (priv. comm.). 
K13: Kim et al. (\cite{kim13}). 
L97: Lewis (\cite{lewis97}). 
LM88: Little-Marenin et al. (\cite{little-marenin88}). 
LM93: Little-Marenin et al. (\cite{little-marenin93}). 
N87: Nakada et al. (\cite{nakada87}). 
N88: Nakada et al. (\cite{nakada88}). 
OB08: Ohnaka \& Boboltz (\cite{ohnaka08euand}). 
SH08: Shintani et al. (\cite{shintani08}). 
SZ06: Szczerba et al. (\cite{szczerba06}). 
$^{\dagger}$: General Catalog of Galactic Carbon Stars by C.B. 
Stephenson (third edition), Alksnis et al. (\cite{alksnis01}).
}
\vspace*{-5mm}

\begin{tabular}{l c c c c c l}\hline
Name & CGCS$^{\dagger}$ & RA  & DEC &  Date       & Det. &
Previous \HOH\ maser detection (references) \\ 
     &      & (J2000) & (J2000) &    2007        &  & \\
\hline
IRAS06017+1011 & 1158 & 06:04:31.4 & $+$10:10:55 & Aug 19 & Y  & Y (E94)\\
BM Gem  & 1653 & 07:20:59.0 & $+$24:59:58        & Aug 19 & N  & N (N87, LM88, EL94, BLM96, L97)\\
IRAS07221-0431& 1698 & 07:24:39.2 & $-$04:37:55 &  Aug 19 &  N  & Y (E94)\\
IRAS08002-3803& 2011 & 08:02:05.1 & $-$38:11:52 &  Aug 19 & N   & N (N88, D89)\\
IRAS18006-3213& 3935 & 18:03:53.0 & $-$32:13:00 &  Aug 20 & N   & N (N88, D89)\\
NC83 & 4222 & 19:15:01.3 & $+$54:17:26 & Aug 20 & {\bf Y} & N 
(LM88, EL94, BLM96)\\
HD189605 & 4595 & 20:01:05.2 & $-$07:21:51 & Aug 20      & N   & N (E94)\\
V778 Cyg  & 4923 & 20:36:07.4 & $+$60:05:26 & Aug 20 & Y       & 
Y (N87, D88, LM88, LM93, EL94, C00, SZ06, SH08, K13, EPC\\
V1415 Cyg & 5548 & 22:01:17.6 & $+$54:32:34 & Aug 20 & Y         &  Y (E94)\\
EU And  & 5848 & 23:19:58.2  & $+$47:14:28 & Aug 20 & Y         & Y
(BLM87, LM88, LM93, EL94, BLM96, C00, OB08, SH08, K13\\
\hline

\label{obslog}
\vspace*{-7mm}

\end{tabular}
\end{center}
\end{table*}

\section{Introduction}
\label{sect_intro}

Silicate carbon stars are characterized by oxygen-rich circumstellar material 
in spite of their carbon-rich photospheres 
(Little-Marenin \cite{little-marenin86}; 
Willems \& de Jong \cite{willems86}).  
As summarized in Ohnaka et al. (\cite{ohnaka06} and references 
therein), a currently believed hypothesis suggests that 
silicate carbon stars have a low-luminosity companion 
(a main-sequence star or a white dwarf) 
and that oxygen-rich material shed by the mass loss in the past, when the 
primary star was an oxygen-rich giant, 
is stored in a circumbinary disk or in a circumstellar disk 
around the unseen companion until the primary star becomes 
a carbon star (Morris \cite{morris87}; Lloyd-Evans \cite{lloyd-evans90}; 
Yamamura et al. \cite{yamamura00}).  
However, the formation mechanisms of such circumbinary or 
circum-companion disks and their relation to binary parameters 
are unknown.  
Moreover, this scenario does not explain the following peculiar chemical 
composition of the photosphere of silicate carbon stars: 
highly enriched in $^{13}$C with $^{12}$C/$^{13}$C$\approx$4--5, in marked 
contrast to $^{12}$C/$^{13}$C $\ga$20 in normal carbon stars 
(Ohnaka \& Tsuji \cite{ohnaka99}).  
This anomalous chemical composition is difficult to 
explain by standard stellar evolution theory and may be related to binarity 
(e.g., Zhang \& Jeffery \cite{zhang13}; Sengupta et al. \cite{sengupta13}). 
Thus, silicate carbon stars represent interesting cases that allow us 
to study the possible effects of binarity on stellar evolution and the mass loss 
accompanied by the formation of disks.

High spatial resolution mid-IR (8--13~\micron) observations with 
the mid-IR interferometric instrument MIDI at the ESO's 
Very Large Telescope Interferometer (VLTI) have spatially 
resolved the dusty environment of silicate carbon stars 
and suggest the presence of circumbinary disks 
(Ohnaka et al. \cite{ohnaka06}; Deroo et al. \cite{deroo07}).  
On the other hand, our MIDI observations of the silicate carbon star 
BM~Gem with a spatial resolution of 20~mas suggest the presence of a 
circum-companion disk (Ohnaka et al. \cite{ohnaka08bmgem}). 
Furthermore, Izumiura et al. (\cite{izumiura08}) detected significant 
emission shortward of $\sim$4000~\AA\ and Balmer lines with the P-Cyg 
profile in BM~Gem, which strongly indicate an accretion 
disk around the putative companion.  

Our 22~GHz water maser mapping of the silicate carbon star EU~And with 
the Very Long Baseline Array (VLBA) 
shows that the masers are aligned with a slightly S-shaped structure 
along a straight line (Ohnaka \& Boboltz \cite{ohnaka08euand}). 
This is similar to what is found in the silicate carbon star V778~Cyg 
by Szczerba et al. (\cite{szczerba06}) using MERLIN and by Engels 
(priv. comm.) using VLBA.  
Ohnaka \& Boboltz (\cite{ohnaka08euand}), as well as Szczerba et al. 
(\cite{szczerba06}) suggest that the masers likely originate in warped 
circum-companion disks viewed almost edge-on. 

The maser imaging of more silicate carbon stars is indispensable for 
understanding the spatial structure of the oxygen-rich circumstellar 
material. 
In order to select objects appropriate for the detailed imaging with 
VLBA, we carried out a small survey of the \HOH\ masers at 22~GHz 
toward 10 silicate carbons stars 
with the Karl G. Jansky Very Large Array (VLA).  Our goal is to detect 
potential 
targets for future VLBA imaging with much better sensitivity than the 
previous observations in the literature.

\begin{figure*}
\resizebox{\hsize}{!}{\rotatebox{-90}{\includegraphics{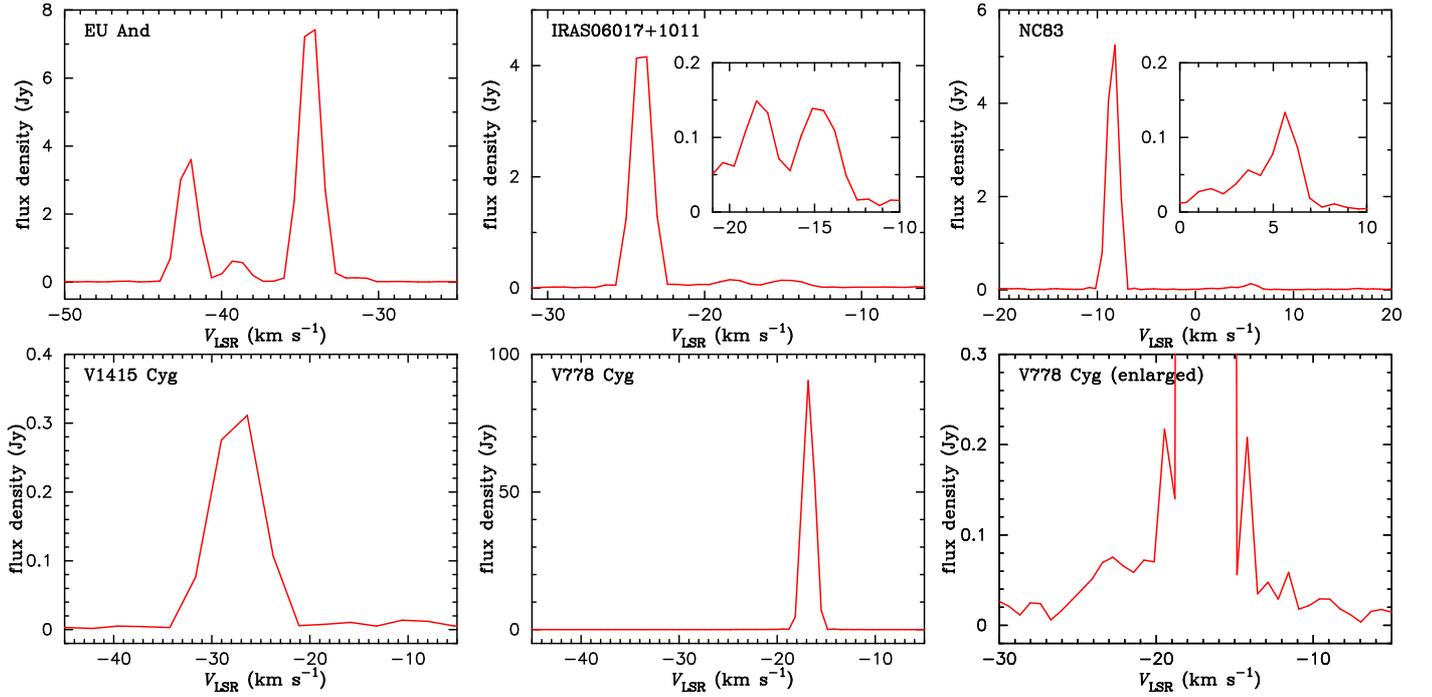}}}
\caption{VLA spectra (vector-averaged cross-power spectra over all 
baselines) of five silicate carbon stars with the detection of 
the \HOH\ masers.  
}
\label{obsspec}
\end{figure*}

\section{Observations and data reduction}
\label{sect_obs}

Table~\ref{obslog} gives a list of our 10 program stars, which were 
selected from the list of the confirmed silicate carbon stars 
in Engels (\cite{engels94}).  
From Table~3 of his paper, which consists of 13 stars, 
we excluded two stars, MC79-11 and 
CGCS3922.  MC79-11 is too southern to observe from VLA.  
The identification of CGCS3922 as a silicate carbon star is not 
entirely established, as Lloyd-Evans (\cite{lloyd-evans91}) notes. 
Besides, the star is 
included in a list of Li-rich ``K giants'' in De la Reza et al. 
(\cite{delareza97}).  
We observed the remaining 11 stars, but the results 
of IRAS07204-1032 (CGCS1682) will be presented in a separate 
paper together with VLBA imaging observations (Boboltz et al., in prep).  
The \HOH\ masers were 
previously detected toward five objects, as summarized 
in Table~\ref{obslog}.  

Observations of the $6_{16}$--$5_{23}$, 22.2-GHz transition of 
\HOH\ were performed using the VLA in A-configuration.  
The VLA is maintained and operated by the National Radio Astronomy 
Observatory (NRAO)\footnote{The National Radio Astronomy Observatory is 
a facility of the National Science Foundation operated under cooperative 
agreement by Associated Universities, Inc.}.  
We observed the 10 silicate carbon stars along 
with their respective extragalactic phase calibrators on 2007 August 19 and 20. 
The program stars except for V1415~Cyg and IRAS08002-3803 
were observed using a bandwidth of 6.25~MHz centered on a line rest frequency 
of 22.23508~GHz.   The VLA correlator produced 128 
spectral channels with a channel spacing of 48.8~kHz ($\sim$0.66~\KMS). 
For V1415~Cyg and IRAS08002-3803, observations 
were conducted with a wider band (12.5 MHz) providing a channel spacing of 
195.3~kHz ($\sim$2.64~\KMS) over 64 channels.  
The central velocity, as well as the velocity range searched 
for our program stars are given in Table~\ref{res_table}.  
The central velocities were set based on the previous maser observations 
or the radial velocity measurements from optical spectral lines.  
However, no information about the radial velocity of IRAS08002-3803 is 
available.  
Therefore, we set the central velocity to zero for this source.

The data were reduced using the standard routines within NRAO's Astronomical 
Image Processing System (AIPS). The absolute flux density scale was established 
using the calibration sources 3C286 and 3C48 assuming 22.2-GHz flux densities of 
$2.59$~Jy and of $1.13$~Jy, respectively.  For each target star, a nearby 
(within 3$^{\circ}$) extragalactic calibrator was observed in order to estimate 
the instrumental and atmospheric phase fluctuations. Observations of these 
calibrators were interspersed with the target source scans at roughly 10 minute 
intervals.  Phase corrections estimated from the calibration sources were 
applied to the target source data.  For each source, an iterative 
self-calibration and imaging procedure was performed to map a single strong 
reference spectral feature in the 22.2-GHz \HOH\ transition.  The resulting 
phase solutions were applied to all channels in the band.  To eliminate 
ringing in channels adjacent to strong maser features, the data were Hanning 
smoothed prior to the production of the final spectra. 
The RMS noise level was estimated from the far wings of the spectra.  
We took the $3\sigma$ RMS noise level as an upper limit on the maser emission 
for non-detection sources. 
While we obtained a data cube of images for each star, the 
spatial resolution of our VLA observations of 87~mas is too low 
to spatially resolve the distribution of the maser spots.  
Therefore, we only present the resulting spectra of our program stars.  

Our program stars are classified as silicate carbon stars 
in the literature based on the IRAS LRS.  
However, for IRAS06017+1011, IRAS07221-0431, NC83, HD189605, and V1415~Cyg, 
the quality of the IRAS LRS is too poor to detect the 10~\micron\ 
silicate emission definitively or study its spectral shape in detail.  
Therefore, we searched for newer mid-IR spectra for these objects.  
IRAS06017+1011 was observed with VLTI/MIDI on 2005 December 22 
(Program ID: 076.D-0250, P.I.: K.~Ohnaka).  
The data were reduced with the MIA+EWS 
ver2.0\footnote{http://home.strw.leidenuniv.nl/\~{}jaffe/ews/index.html}, 
and the spectra were calibrated using HD39400 (K1.5IIb) as a spectroscopic 
standard star, using the method described in Ohnaka et al. (\cite{ohnaka06}).  
NC83 and HD189605 were observed with the InfraRed Spectrometer (IRS, 
Houck et al. \cite{houck04}) onboard the Spitzer Space Telescope 
(Werner et al. \cite{werner04}) on 
2004 December 9 and 2005 October 11, respectively, using the Short-High 
and Long-High modes with a spectral resolution of $\sim$600 
(Program ID: P03235, P.I.: C.~Waelkens).  
We downloaded the Post Basic Calibrated Data (PBCD) from the 
Spitzer Heritage Archive.  
As presented in Sect.~\ref{sect_res}, the new mid-IR spectra confirm 
the silicate emission in these stars.

\begin{table*}
\begin{center}
\caption {Result of the VLA \HOH\ maser observations of our program stars.  
The central velocity and the searched velocity range, as well as 
the RMS noise level are listed. 
For positive detections, the velocity ($V_{\rm peak}$) and intensity 
($S_{\rm peak}$) of each peak are given, together with the integrated flux 
($S_{\rm I}$).  
For non-detection sources, the upper limit (3$\sigma$ RMS noise) is given. 
}
\vspace*{-5mm}

\begin{tabular}{l c c c c c c c c}\hline
Name & CGCS$^{\dagger}$ & \VLSR\ & Searched \VLSR\ & Maser \VLSR\ &
$V_{\rm peak}$  & $S_{\rm peak}$ & $S_{\rm I}$ & RMS noise  \\
     &                &        & range    (\KMS)  & range (\KMS)                 &
(\KMS)        & (Jy)       & ($10^{-21}$~W~m$^{-2}$) & (mJy) \\
\hline
IRAS06017+1011 & 1158 & $-13.0$ & $-54.6$ ... $+28.3$ & $-24.0$ ... $-15.1$ & $-24.0$ & 4.2 & 6.5 & 5.3\\
 &  &  &  &  & $-18.4$ & 0.15 & & \\
 &  &  &  &  & $-15.1$ & 0.14 & & \\
BM Gem  & 1653 & $+76.1$ & $+34.0$ ... $+118.2$ & --- & --- & $<$0.015 & --- & 
5.1\\
IRAS07221-0431& 1698 & $+79.0$ & $+36.9$ ... $+121.1$ & --- & --- & $<$0.023 &
--- & 7.8\\
IRAS08002-3803& 2011 & 0.0 & $-42.1$ ... $+42.1$ & --- & --- & $<$0.045 & --- &
15.0\\
IRAS18006-3213& 3935 & $+176.7$ & $+134.6$ ... $+218.8$ & --- & --- & $<$0.033 &
--- & 11.0\\
NC83 & 4222 & $-6.7$ & $-48.3$ ... $+34.6$ & $-8.2$ ... $+5.6$ & $-8.2$ & 5.2
& 6.9 & 7.3\\
 &  &  &  &  & $+1.7$ & 0.03 & & \\
 &  &  &  &  & $+3.6$ & 0.06 & & \\
 &  &  &  &  & $+5.6$ & 0.13 & & \\
HD189605 & 4595 & $+58.2$ & $+16.1$ ... $+100.3$ & --- & --- & $<$0.024 & --- &
7.9\\
V778 Cyg  & 4923 & $-17.0$ & $-58.3$ ... $+24.6$ & $-25.0$ ... $-14.2$
&$-25.0$...$-20.1$ & 0.05--0.07 & 100 & 22.0\\
 & &  &  &  & $-19.5$ & 0.22 & & \\
 & &  &  &  & $-16.8$ & 90.4 & & \\
 & &  &  &  & $-14.2$ & 0.21 & & \\
V1415 Cyg & 5548 & 0.0 & $-81.7$ ... $+81.6$ & $-34.3$ ... $-21.1$ & $-26.4$ & 0.31
& 2.1 & 3.5\\
EU And  & 5848 & $-36.0$ & $-77.5$ ... $+5.4$ & $-42.0$ ... $-34.1$ & $-42.0$
& 3.6 & 16 & 5.2\\
 & &  &  &  & $-39.3$ & 0.62 & & \\
 & &  &  &  & $-34.1$ & 7.4  & & \\
\hline

\label{res_table}
\vspace*{-7mm}

\end{tabular}
\end{center}
\end{table*}

\section{Results and discussion}
\label{sect_res}

We detected the \HOH\ maser emission from 5 out of 10 sources: 
EU~And, IRAS06017+1011, NC83, V778~Cyg, and V1415~Cyg, with NC83 
being the first maser detection.  
Figure~\ref{obsspec} shows the \HOH\ maser spectra of these 
five stars.  
The velocity range of the detected maser peaks, the velocity and intensity 
of each peak, and the integrated flux are given in Table~\ref{res_table}. 
For the non-detection sources, we give the upper limit of the maser 
intensity set by the 3$\sigma$ RMS noise estimated as described in 
Sect.~\ref{sect_obs}. 

For NC83, Little-Marenin et al. (\cite{little-marenin88}), 
Engels \& Leinert (\cite{engels&leinert94}), and Benson \& Little-Marenin 
(\cite{benson96}) searched for \HOH\ masers 
at more than 10 epochs in total in the 1980s and 1990s but detected no \HOH\ 
masers.  In 2007 we detected a strong peak with 
4.1~Jy, which would have been detectable in the previous observations.  
For example, the upper limit of 0.1~Jy reported by Engels \& Leinert 
(\cite{engels&leinert94}) means a increase in the maser intensity 
by a factor of at least 40.  
On the other hand, IRAS07221-0431 showed a single-peaked \HOH\ maser with 
0.75~Jy at \VLSR\ = 79~\KMS\ in 1992 (Engels \cite{engels94}), but 
we detected no \HOH\ masers, despite the detection limit (3$\sigma$ RMS) 
of 0.02~Jy. 
This means that the maser intensity has decreased by more than a factor of 
38. 
While significant time variations in the \HOH\ masers toward silicate carbon 
stars are not unusual, the drastic variation by a factor of $\sim$40 seen 
in NC83 and IRAS07221-0431 is unique.  This might be caused by some mechanism 
different from the usual erratic variations in the masers in stars such as 
V778~Cyg and IRAS06017+1011 (as described below).  NC83 and IRAS07221-0431 
are worth long-term monitoring.  

In addition to this ``on/off'' variation, we also detected noticeable 
time variations in the maser intensity and spectral shape in IRAS06017+1011 
and V778~Cyg.  
Engels (\cite{engels94}) detected a strong single-peaked \HOH\ 
maser (20~Jy) at $-13$~\KMS\ toward IRAS06017+1011 in 1992, 
while our VLA spectrum obtained 15 years later is very different, showing a 
peak (4.2~Jy) at $-24$~\KMS\ and much weaker peaks of 150 and 
140~mJy at $-18$ and $-15$~\KMS, respectively. 
Our VLA spectrum of V778~Cyg is also remarkably different from the MERLIN 
spectrum taken in 2001 (Szczerba et al. \cite{szczerba06}), which shows 
the strongest peak at $-15$~\KMS\ and a much weaker peak at $-17$~\KMS.

The VLA \HOH\ maser spectra of EU~And, IRAS06017+1011, and NC83 are 
characterized 
by blue- and red-peaks separated by 8, 9, and 14~\KMS, respectively, 
and weaker peaks in between.  
The VLA spectrum of V778~Cyg is dominated by a single peak at $-17$~\KMS, 
but there are weaker peaks of 0.22~Jy at 
$-19.5$~\KMS\ and 0.21~Jy at $-14.2$~\KMS\ on each side of the primary peak 
(Fig.~\ref{obsspec}, bottom right).  
In addition, there is some low-level emission 
from $-24.0$ to $-20.1$~\KMS\ that is 3-4$\sigma$ above the 
RMS noise of 0.02~Jy.  
This means a velocity separation of $\sim$10~\KMS\ between the most 
blue- and red-shifted features.  
The \HOH\ maser spectrum of V1415~Cyg is also dominated by a single peak.  
This spectral feature is much broader ($\sim$10~\KMS) than the features
in our other spectra, but the spectrum of V1415~Cyg was 
recorded with a much wider channel 
spacing of $\sim$2.64~\KMS.  Therefore, this single peak is likely a blend 
of several narrower features.  In this case, the velocity width of this 
single peak of $\sim$10~\KMS\ would approximately correspond to the velocity 
separation between the most blue- and red-shifted peaks.  

\begin{figure*}
\resizebox{\hsize}{!}{\rotatebox{-90}{\includegraphics{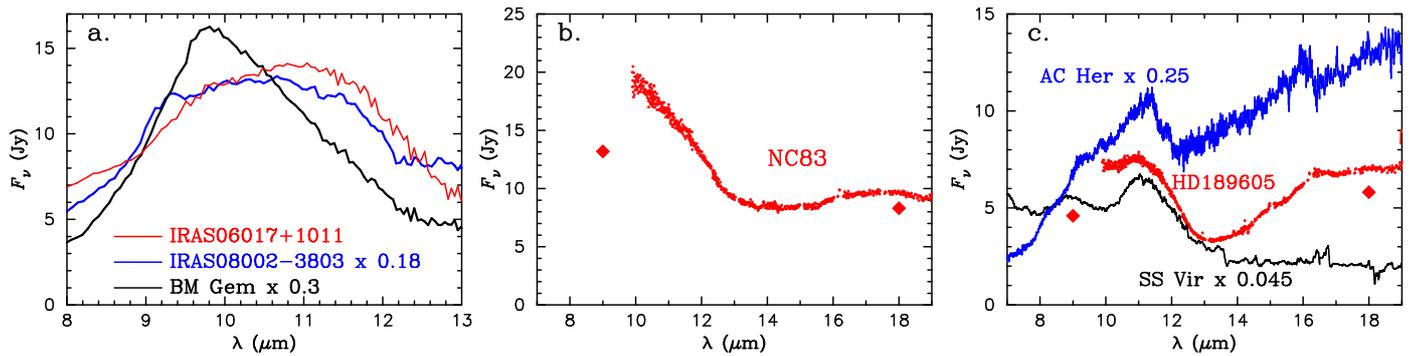}}}
\caption{New mid-IR spectra of IRAS06017+1011 obtained with VLTI/MIDI 
({\bf a}), NC83 ({\bf b}), and HD189605 ({\bf c}), both obtained with 
Spitzer/IRS.  
Also plotted are the scaled spectra of IRAS08002-3803, BM~Gem, AC~Her,  
and SS~Vir.    
}
\label{midir_spec}
\end{figure*}

The measured velocity separations can be used to estimate the 
projected rotational velocity of the maser emitting region in circumbinary or 
circum-companion disks.  
We suggest that the masers in IRAS06017+1011, NC83, and V1415~Cyg 
originate in circum-companion disks similar to EU~And and V778~Cyg 
(Ohnaka \& Boboltz \cite{ohnaka08euand}; Szczerba et al. 
\cite{szczerba06}) for the following reason.  
The amount of the IR excesses toward these stars is similar and modest 
(see Table~3 of Engels \cite{engels94}) and, therefore, implies that 
the oxygen-rich circumstellar material is optically thin.  
As Yamamura et al. (\cite{yamamura00}) argue, such material cannot 
exist stably in a circumbinary or circum-primary disk because the oxygen-rich 
material would be blown away by the intense radiation pressure from the 
primary carbon star.  

It should be noted that 
the velocity separation measured at a given epoch can be narrower 
than the true velocity separation due to significant time variations in the 
maser intensity and the spectral shape as described above.  
Therefore, 
we checked the \HOH\ maser spectra previously presented in the literature 
(references listed in Table~\ref{obslog}) to estimate the true velocity 
separation for the sources with the positive maser detection with VLA 
except for NC83 (our VLA observation is the first maser detection for this 
source).  
The maximum maser velocity range obtained from the past and present 
maser spectra is 
$-42.6$...$-29.5$~\KMS, $-24.0$...$-13.0$~\KMS, $-24.0$...$-14.2$~\KMS, 
for EU~And, IRAS06017+1011, and V778~Cyg, 
respectively\footnote{Little-Marenin et al. (\cite{little-marenin88}) report 
a possible detection of a peak at $-12.4$~\KMS\ toward EU~And.  
However, because it is 
not a definitive detection, we did not include this peak.}.  
Therefore, 
the velocity separation in EU~And, IRAS06017+1011, V778~Cyg, and NC83, 
are 13.1, 11.0, 9.8, 13.8~\KMS, respectively.  
In the case of V1415~Cyg, the only previous detection is reported by 
Engels (\cite{engels94}), who detected a narrow, single peak at $-28.2$~\KMS. 
As mentioned above, the single peak in our VLA spectrum, which appears 
at roughly the same velocity, is much broader.  Therefore, we took the width 
of this broad feature ($\sim$10~\KMS) as the velocity separation.

A half of measured velocity separations corresponds to 
the projected rotational velocity of the putative circum-companion disks.  
Therefore, the observed velocity separations of 10--14~\KMS\ translate 
into the projected rotational velocities of 5--7~\KMS\ for five stars 
shown in Fig.~\ref{obsspec}. 
If we adopt 0.5--0.8~\MSOL\ for the companion mass as in EU~And 
(Ohnaka \& Boboltz \cite{ohnaka08euand}) and assume the Keplerian 
rotation, the radius of the maser emitting region is estimated to be 
10--32~AU.  Adopting a higher mass of 1.7~\MSOL\ derived for V778~Cyg 
by Babkovskaia et al. (\cite{babkovskaia06}) leads to a radius of 68~AU 
for the rotational velocity of 5~\KMS.  
Because the velocity separations estimated from the past and present maser 
spectra are still a lower limit, the estimated projected rotational 
velocities are a 
lower limit.  The projected rotational velocities themselves give a lower 
limit on the (de-projected) rotational velocities.  
Therefore, the radius of the maser emitting region estimated above is an 
upper limit.

If the masers originate in 
circum-companion disks, the velocity range of the maser spectra is expected 
to systematically drift due to the orbital motion of the companion 
around the primary star if observed long enough.  
However, this is difficult to detect for two reasons.  
Firstly, the orbital period 
is expected to be as long as a few hundred years.  For example, 
the modeling of the masers toward V778~Cyg by Babkovskaia et al. 
(\cite{babkovskaia06}) suggests a mass of 1~\MSOL\ and 1.7~\MSOL\ for 
the primary star and the companion, respectively, with a separation of 
80~AU.  This translates into an orbital velocity of $\sim$2~\KMS\ and 
a period of $\sim$440~years.  Secondly, the small systematic drift in the 
velocity range caused by the orbital motion is masked by the erratic 
time variation of the masers as mentioned above.  
We searched for possible systematic drifts in the velocity range of the 
masers in the spectra of V778~Cyg and EU~And, which are the best targets 
for this purpose, thanks to the ample maser observations in the past 
(see references in Table~\ref{obslog}), 
but could find no definitive systematic drift.

The new mid-IR spectra of IRAS06017+1011, NC83, and HD189605, 
shown in Fig.~\ref{midir_spec}, reveal the silicate emission, 
confirming that these stars are silicate carbon stars.  
While the Spitzer/IRS spectra of NC83 and HD189605 are 
available only longward of 10~\micron, the combination with the 9~\micron\  
flux measured by AKARI (Ishihara et al. \cite{ishihara10}) indicates the 
10~\micron\ silicate feature.  
The spectra of IRAS06017+1011 and HD189605 exhibit a bump 
centered at 11--11.5~\micron, which is clear when compared with BM~Gem 
in Fig.~\ref{midir_spec}a.  This bump is also seen in the silicate carbon 
stars IRAS08002-3803 (Ohnaka et al. \cite{ohnaka06}, also plotted in 
Fig.~\ref{midir_spec}a) and IRAS18006-3213 
(Deroo et al. \cite{deroo07}). 
A comparison with the spectrum of the carbon-rich Mira SS~Vir obtained 
with the Infrared Space Observatory (ISO), shown in Fig.~\ref{midir_spec}c, 
suggests that the bump centered at 11~\micron\ resembles the SiC feature 
often observed in usual carbon stars.  
Therefore, 
we suggest that the bump at 11--11.5~\micron\ is due to SiC, which originates 
in the ongoing mass loss from the carbon-rich primary star.  

It is noteworthy that the Spitzer/IRS spectrum of HD189605 shows no 
signatures of crystalline silicate, although the spectral resolution 
of 600 is sufficient to resolve its sharp features.  
In Fig.~\ref{midir_spec}c, 
the ISO spectrum of the post-AGB star AC~Her obtained with a spectral 
resolution of 750 is plotted.  
AC~Her, one of the best examples, shows fine, sharp crystalline 
silicate features, which can be seen, for example, at 
11.4, 16, and 18~\micron\ (Molster et al. \cite{molster02}).  
None of these features are seen in HD189605.

Furthermore, HD189605 was detected in the near-UV with the 
Galaxy Evolution Explorer (GALEX) with a flux of 5.1--7.7~$\mu$Jy at 
2271~\AA \footnote{None of the other silicate carbon stars in our list 
is found in the GALEX catalog at http://galex.stsci.edu/GR6/}. 
The offset of the object detected 
with GALEX (GALEX J200105.1-072151) from the position of HD189605 measured 
in the optical and IR is 0\farcs2--0\farcs9, which is within 
the positional error of the GALEX data.  
This means that the near-UV emission is likely associated with HD189605.  
The UV emission may originate from an accretion disk around the putative 
companion as in BM~Gem (Izumiura et al. \cite{izumiura08}). 
Therefore, HD189605 is a good candidate for studying the accretion process 
in silicate carbon stars.  

There seems to be no clear correlation between the 
maser detection and the amount of the IR excess.  For example, 
IRAS06017+1011 and BM~Gem show approximately the same amount of the 
IR excess (see their SEDs in Figs.~3b and 3c in Kwok \& Chan
\cite{kwok93}).  However, IRAS06017+1011 shows masers, while 
BM~Gem has not shown any masers so far.  
The detection of the masers does not show a clear correlation 
with the shape of the silicate feature, either.  
For example, IRAS06017+1011 shows a broad 
silicate feature with a 11.5~\micron\ bump, while EU~And 
shows a narrow silicate feature.  Nevertheless, both objects show maser 
emission.  This can be understood if SiC in the ongoing mass loss 
from the carbon-rich primary star is responsible for the 
11.5~\micron\ bump.  In this case, the shape of the 10~\micron\ 
feature is not related to the oxygen-rich material.

\section{Conclusions}
\label{sect_concl}

We presented the result of a survey of the 22~GHz \HOH\ masers toward 10 
silicate carbon stars using the VLA.  \HOH\ masers were detected in five 
stars, including NC83, for which masers have been detected for the first time. 
The \HOH\ maser spectra show a velocity separation of 10--14~\KMS\ 
between the most blue- and red-shifted peaks.  
This suggests a lower limit of the rotation velocity of 
the circum-companion disks of 5--7~\KMS\ and 
an upper limit of the radius of the maser emitting region of 
10--68~AU for a companion mass of 0.5--1.7~\MSOL.  

We confirmed the 10~\micron\ silicate feature in 
IRAS06017+1011, NC83, and HD189605 using newer mid-IR spectra obtained 
with VLTI/MIDI and the Spitzer Space Telescope.  
The silicate feature in IRAS06017+1011 and HD189605 shows a broad bump 
centered at 11--11.5~\micron.  
We suggest that SiC in the ongoing mass loss from the carbon-rich 
primary star, not crystalline silicate, may be responsible for this bump.  
We also detected near-UV emission toward HD189605 with GALEX.  

The radius of the maser emitting region of 10--68~AU translates into 
5--68~mas for distances of 1--2~kpc.  This is resolvable with 
VLBA.  High spatial resolution imaging of more silicate carbon stars with VLBA 
is feasible and necessary to reveal the spatial distribution of 
oxygen-rich material.

\end{document}